# AETHER THEORY AND THE PRINCIPLE OF RELATIVITY[1]

## Joseph Levy

4 Square Anatole France, 91250 St Germain-lès-Corbeil, France
E. Mail: levy.joseph@orange.fr

## ABSTRACT

This paper completes and comments on some aspects of our previous publications. In ref [1], we have derived a set of space-time transformations referred to as the extended space-time transformations. These transformations, which assume the existence of a preferred aether frame and the variability of the one-way speed of light in the other frames, are compared to the Lorentz-Poincaré transformations. We demonstrate that the extended transformations can be converted into a set of equations that have a similar mathematical form to the Lorentz-Poincaré transformations, but which differ from them in that they connect reference frames whose co-ordinates are altered by the systematic unavoidable measurement distortions due to length contraction and clock retardation and by the usual synchronization procedures, a fact that the conventional approaches of relativity do not show. As a result, we confirm that the relativity principle is not a fundamental principle of physics [i.e, it does not rigorously apply in the physical world when the true co-ordinates are used]. It is contingent but seems to apply provided that the distorted coordinates are used. The apparent invariance of the speed of light also results from the measurement distortions. The space-time transformations relating experimental data, therefore, conceal hidden variables which deserved to be disclosed for a deeper understanding of physics.

---

[1] Version supplemented by further explanations published in "Ether space-time and cosmology", volume 1, Michael C. Duffy and Joseph Levy Editors.

# 1. Introduction

Special relativity regards the relativity principle as a fundamental principle of physics [2]. In other words the laws of physics must be identical in all inertial frames[*]. But a fundamental principle supposes that the variables which are involved in the laws are exactly measured in all these frames. If the principle applies exclusively when the measurement brings about a certain distortion of the variables, it loses its character as a fundamental principle of physics. Such a principle must be qualified as contingent. A contingent principle may have a certain practical value, but it does not enable us to directly know the mechanisms which determine the physical processes, and it gives a distorted view of reality. It is therefore justified to try and show the contingent character of a principle when it is suspected and to determine the true co-ordinates which are not altered by the measurement distortions.

A contingent principle is not absolute, it depends on the conventions chosen for the measurements, and if we use other conventions the principle will not apply. It is important to check to what extent these considerations concern the relativity principle. The question will be developed in the following chapters.

Most authors today ignore the existence of a preferred aether frame, despite the experimental and theoretical arguments that have been developed recently [1]. Others regard the relativity principle as a fundamental principle compatible with the existence of a preferred aether frame [3].They treat as inertial the frames associated to moving bodies, provided that they are not submitted to physical influences other than the aether, which implies the equivalence of these frames for the description of the physical laws. (But, as we have seen in ref [1], under the action of the aether drift the said frames cannot be strictly inertial, and although we will use this term which is sanctioned by use, we must be aware that it is an approximation only valid when the aether drift is weak).

In previous publications [1], assuming the existence of a preferred aether frame and the variability of the one-way speed of light in the other frames, we have derived a set of space-time transformations, referred to as the extended space-time transformations, which do not obey in all generality the relativity principle, in that the transformations connecting the aether frame with any other 'inertial frame', possess a mathematical form different from the transformations connecting any pair of 'inertial frames'.

In the following chapters, we shall demonstrate that these transformations can be converted into transformations which assume the same mathematical form as the Lorentz-Poincaré transformations, but which fundamentally differ from them, in that the co-ordinates of the reference frames they connect, (except for the preferred frame), are shown to be altered by systematic measurement distortions[**]. In particular, they depend on a questionable synchronization procedure which generates a synchronism discrepancy effect, whose magnitude varies with the pair of 'inertial frames' considered, and if we change the synchronization procedure, these transformations will not apply. Therefore, only the laws of physics relating distorted variables will be invariant and not the true laws.

---

[*] Poincaré gives the principle a different formulation. According to Poincaré's relativity principle, it would be impossible by means of an experiment internal to a given inertial frame to highlight the absolute motion of this frame.

As an example of this fact, let us revisit the case of the two rockets receding from one another along a same line in a frame S different from the aether frame. At instant $t_0$, the rockets meet at a point O, and then they continue on their way, symmetrically, at speed $v$, towards two points A and B placed at equal distance from point O. We first suppose that the speeds of the rockets are exactly measured. At the instant they meet, the clocks inside the rockets are set to $t_0$. Of course the rockets have different speeds with respect to the aether frame and, therefore, due to clock retardation, their clocks will display different readings when they reach points A and B in contradiction with the relativity principle. (Only if frame S were at rest with respect to the aether frame, would they display the same reading).

Yet if the speeds of the rockets are measured with clocks placed at A, O and B synchronized by means of the Einstein-Poincaré procedure with light signals (or by slow clock transport), the clocks inside the rockets will display the same reading when they reach points A and B, a fact which seems in agreement with the relativity principle. This result follows from the systematic error made in measuring the speeds, when, using the Einstein-Poincaré procedure, one assumes the isotropy of the one way speed of light.

But one cannot conclude that the relativity principle is a fundamental principle of physics in the physical world, if it depends on a synchronization procedure which gives rise to a systematic measurement error.

---

[**] In ref [1B], we have applied the term Lorentz-Poincaré transformations to the transformations which assume the same mathematical form as the conventional transformations. In fact, as we shall see the term should be reserved more specifically for the transformations which connect any 'inertial system' to the fundamental frame in which the space and time co-ordinates are not altered by measurement distortions (which is not the case in the usual applications).

After a brief reminder of the extended space-time transformations, we shall compare them to the Lorentz-Poincaré transformations, making a distinction between Einstein's approach, which denies the existence of a preferred aether frame, and Poincaré's approach, which assumes such an aether frame. We shall verify that our approach departs completely from those we have just mentioned, in that it starts from the Galilean transformations, and demonstrates that the co-ordinates generally used in the space-time transformations result from the distortions affecting the Galilean co-ordinates, which are in fact the true co-ordinates. Although this fact is not recognized by the conventional approaches of relativity theory, it concerns all the measurements carried out in the Earth frame.

## 2. Brief reminder of the extended space-time transformations

We start from the space-time transformations connecting any pair of 'inertial frames' which are not submitted to measurement distortions. These Galilean transformations are not those which are determined experimentally because the distortions cannot totally be avoided when the experiments are performed (such distortions result from length contraction, clock retardation and arbitrary clock synchronization [4]). Submitting the Galilean transformations to these distortions, we obtain the extended space-time transformations which enable us to show how the distortions act. Finally these transformations will be given another mathematical form that will permit us to easily compare our approach to the conventional approaches of special relativity.
(If the reader is already familiar with the subject, he may skip this paragraph).
Consider to this end three co-ordinate systems $S_0$, $S_1$ and $S_2$ (Fig 1). $S_0$ is at rest in the Cosmic substratum (aether frame), $S_1$ and $S_2$ are moving along the common $x$-axis with rectilinear uniform motion. We propose to derive the space-time transformations connecting the co-ordinate systems $S_1$ and $S_2$.
At the initial instant, the origins of the three co-ordinate systems O, O' and O'' are coincident. At this instant a vehicle coming from the $-x_2$ region passes by O'' and then continues on its way with rectilinear uniform motion along a rigid path AB toward point B. We shall refer to the speeds between $S_i$ and $S_j$ as $v_{ij}$ and to the speed of the vehicle with respect to $S_0$ as $V$ (with $V > v_{02}$). The line AB which is firmly fixed to the system $S_2$ and is aligned along the $x_2$-axis would assume the length $\ell_0$ if it was at rest in $S_0$, but as a result of its motion its length is reduced to $\ell = \ell_0 \sqrt{1 - v_{02}^2/C^2}$. The origin A of the path permanently coincides with O''.

When the vehicle reaches point B, it meets a clock equipped with a mirror firmly fixed to the system $S_1$ and standing at a point B' of this system (so that when the vehicle arrives at point B, B and B' are coincident (figure 1).

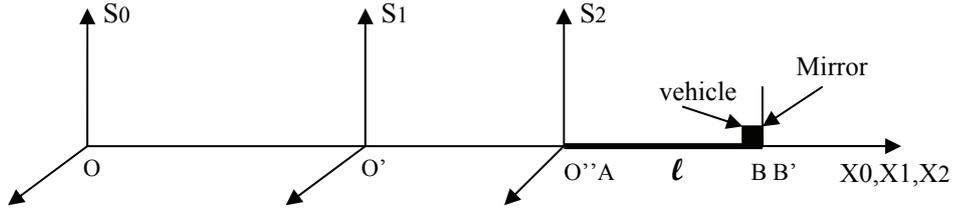

Fig1. When the vehicle reaches point B, it meets a clock equipped with a mirror firmly fixed to the system $S_1$ at a point B' of this system.

Let us determine the distance and the time needed by the vehicle to reach point B from the point of view of an observer at rest with respect to the co-ordinate system $S_1$. We shall first determine the true co-ordinates.
When the vehicle has covered the distance $\ell$ relative to point O', the co-ordinate system $S_2$ has moved with respect to the system $S_1$ a distance equal to:

$$\frac{v_{12}}{V - v_{01}} \ell$$

When the vehicle has covered this distance in its turn, $S_2$ has moved an additional distance equal to:

$$\frac{v_{12}}{V - v_{01}} \left( \frac{v_{12}}{V - v_{01}} \right) \ell = \frac{v_{12}^2}{(V - v_{01})^2} \ell$$

And so on. Therefore, in order to reach point B, the vehicle must cover with respect to $S_1$ a distance $X_{1r}$ equal to:

$$\ell_0 \sqrt{1 - v_{02}^2/C^2} \left( 1 + \frac{v_{12}}{V - v_{01}} + \frac{v_{12}^2}{(V - v_{01})^2} + \ldots + \frac{v_{12}^n}{(V - v_{01})^n} + \ldots + \ldots \right)$$

Thus: $X_{1r} = \ell_0 \sqrt{1 - v_{02}^2/C^2} \dfrac{V - v_{01}}{V - v_{01} - v_{12}} = \ell_0 \sqrt{1 - v_{02}^2/C^2} \dfrac{V - v_{01}}{V - v_{02}}$ (1)

We note that we make use of the Galilean law of composition of velocities. As we shall show, the relativistic law of composition of velocities applies only when the co-ordinates are altered by the systematic measurement distortions.

The distance $X_{1r}$ is the true distance measured with a standard non-contracted by the movement. Using a contracted standard, the observer at rest in $S_1$ will find an apparent distance equal to:

$$X_{1app} = \frac{\ell_0 \sqrt{1 - v_{02}^2/C^2}}{\sqrt{1 - v_{01}^2/C^2}} \frac{V - v_{01}}{V - v_{02}} \tag{2}$$

Now, in order to measure in $S_1$ the time needed by the vehicle to reach point B, we must beforehand synchronize two clocks placed in O' and B'. To this end, we send a light signal from O' to B'. After reflection the signal comes back to O'.

The method of Poincaré-Einstein treats the clock reading $\frac{t_1 + \bar{t}_1}{2} \sqrt{1 - v_{01}^2/C^2}$ as the one-way transit time of light. In reality, it is the *apparent* average transit time of light $\tau_{1app}$. The real transit time of light from O' to B' is in fact:

$$t_1 = \frac{X_{1r}}{C - v_{01}} \tag{3}$$

And from B' to O': $\bar{t}_1 = \frac{X_{1r}}{C + v_{01}}$ (4)

Taking account of clock retardation in $S_1$, the synchronism discrepancy $\Delta$ between the clocks placed at O' and B' is therefore given by: (see ref [1])

$$\Delta = t_1 \sqrt{1 - v_{01}^2/C^2} - \frac{(t_1 + \bar{t}_1)}{2} \sqrt{1 - v_{01}^2/C^2} = \frac{(t_1 - \bar{t}_1)}{2} \sqrt{1 - v_{01}^2/C^2}$$

From (1), (3) and (4) we obtain:

$$\Delta = \frac{v_{01} \ell_0}{C^2} \frac{\sqrt{1 - v_{02}^2/C^2}}{\sqrt{1 - v_{01}^2/C^2}} \frac{V - v_{01}}{V - v_{02}}$$

Now, the true time needed by the vehicle to cover the distance $X_{1r}$ in $S_1$ is:

$$T_{1r} = \frac{X_{1r}}{V - v_{01}} = \frac{\ell_0 \sqrt{1 - v_{02}^2/C^2}}{V - v_{02}} \qquad \text{(from (1))}$$

This time is the universal time that clocks would display if they were at rest in the aether frame (in which there is no speed of light anisotropy and no clock retardation).

But in $S_1$ we must take account of the synchronism discrepancy effect and of clock retardation, so that the experimental *apparent* time obtained when we use the synchronization procedure of Poincaré-Einstein [4] is:

$$T_{1app} = T_{1r} \sqrt{1 - v_{01}^2/C^2} - \Delta$$

$$= \ell_0 \frac{\sqrt{1-v_{02}^2/C^2}}{\sqrt{1-v_{01}^2/C^2}} \frac{(1-v_{01}V/C^2)}{V-v_{02}} \tag{5}$$

From expressions (2) and (5) we obtain

$$V_{1app} = \frac{X_{1app}}{T_{1app}} = \frac{V-v_{01}}{1-v_{01}V/C^2}$$

This expression takes the same form as the composition of velocities law of special relativity but, obviously, it has not the same meaning. It is an apparent speed resulting from the measurement distortions.

Expressions (2) and (5) can be expressed as functions of $T_{2app}$ and $X_{2app}$. We note that the length of the rigid line AB is arbitrary, and since it is measured in S$_2$ with a contracted standard, we have $X_{2app} = \ell_0$

we also note that 
$$X_{2app} = \frac{V-v_{02}}{1-v_{02}V/C^2} T_{2app} \tag{6}$$

replacing $\ell_0$ with $X_{2app}$ in (2), we obtain:

$$X_{1app} = X_{2app} \frac{\sqrt{1-v_{02}^2/C^2}}{\sqrt{1-v_{01}^2/C^2}} \frac{V-v_{01}}{V-v_{02}} \tag{7}$$

and replacing $\ell_0$ with expression (6) in (5) gives:

$$T_{1app} = T_{2app} \frac{\sqrt{1-v_{02}^2/C^2}}{\sqrt{1-v_{01}^2/C^2}} \frac{(1-v_{01}V/C^2)}{(1-v_{02}V/C^2)} \tag{8}$$

Expressions (7) and (8) are the extended space-time transformations.
We can now see that, contrary to Einstein's relativity, $v_{01}$ and $v_{02}$, which are the velocities of S$_1$ and S$_2$ with respect to the aether frame, are systematically omnipresent in the equations.

**3. Relation between the extended space-time transformations and the Lorentz-Poincaré transformations.**

Relativity theory, despite its limitations, has permitted the formulation of several physical laws, and therefore, it appears legitimate to estimate the differences between conventional relativity and more recent approaches, and to measure their implications. Actually the answer is not so easy because there are today different conceptions of relativity. We shall envisage successively the most generally accepted conceptions, Einstein's relativity and Poincaré's theory.

### 3.1. Einstein's relativity [2]

The answer to our question is easier as regards Einstein's relativity, since this approach does not assume the existence of a fundamental reference frame.

As we have seen, when we deal with two 'inertial systems' $S_1$ and $S_2$ receding from the aether frame at speeds $v_{01}$ and $v_{02}$ the extended space-time transformations for a vehicle moving at speed $V$ relative to the aether frame are:

$$X_{2app} = X_{1app} \frac{\sqrt{1-v_{01}^2/C^2}}{\sqrt{1-v_{02}^2/C^2}} \frac{V-v_{02}}{V-v_{01}} \qquad (9)$$

$$T_{2app} = T_{1app} \frac{\sqrt{1-v_{01}^2/C^2}}{\sqrt{1-v_{02}^2/C^2}} \frac{(1-v_{02}V/C^2)}{(1-v_{01}V/C^2)} \qquad (10)$$

Here, the difference with special relativity is obvious because in special relativity theory there is no preferred aether frame, and therefore the speeds $v_{01}$ and $v_{02}$ do not mean anything. Nevertheless, in the specific case where the system $S_1$ is at rest in the Cosmic substratum (aether frame) we have $v_{01} = 0$. Therefore:

$$X_{2app} = X_0 \frac{1-v_{02}/V}{\sqrt{1-v_{02}^2/C^2}} = \frac{X_0 - v_{02}T_0}{\sqrt{1-v_{02}^2/C^2}} \qquad (11)$$

$$T_{2app} = T_0 \frac{1-v_{02}V/C^2}{\sqrt{1-v_{02}^2/C^2}} = \frac{T_0 - v_{02}X_0/C^2}{\sqrt{1-v_{02}^2/C^2}} \qquad (12)$$

We note that $X_0, T_0$ and $v_{02}$ in expressions (11) and (12) are not subjected to measurement alterations. Here, we can see that there is a formal similarity between the extended space-time transformations and Einstein's transformations although their meaning is quite different. The similarity can be extended, for example, to the cases where the speed of the moving bodies under consideration is very fast in comparison with the absolute speed of the Earth frame. (This is the case of elementary particles moving at a speed close to the speed of light). In such cases, the fact that the Earth frame has absolute motion, hardly affects the results of the calculations.

### 3.2. Poincaré's theory [5]

Poincaré's theory assumes the existence of a preferred aether frame and at the same time that all 'inertial frames' are equivalent for the description of the physical laws. For Poincaré, the relativity principle applies without restrictions and, apparently, nothing differentiates the co-ordinates of the aether frame and the co-ordinates of the other frames. Therefore the space-time transformations connecting any pair of 'inertial frames' take the form:

$$x' = \frac{x - vt}{\sqrt{1 - v^2/C^2}} \quad (13) \qquad \text{and} \qquad t' = \frac{t - vx/C^2}{\sqrt{1 - v^2/C^2}} \quad (14)$$

Nothing in Poincaré's approach indicates that the co-ordinates $x$, and $t$ in frame S and $x'$ and $t'$ in frame S' result from the measurement distortions which affect the Galilean co-ordinates.

Our approach differentiates from Poincaré's approach, in that it makes a large difference between the true co-ordinates and the apparent co-ordinates, which are derived from the Galilean co-ordinates by submitting them to the systematic measurement distortions (due to length contraction, clock retardation and arbitrary clock synchronization).

In our approach, the only case where $x$ and $t$ in (13) and (14) are the true co-ordinates, is when the transformations connect any 'inertial frame' with the aether frame. Indeed, from (7) and (8), assuming that $S_1$ is at rest in the Cosmic substratum, we have:

$$X_{2app} = \frac{X_0 - v_{02}T_0}{\sqrt{1 - v_{02}^2/C^2}} \quad (15) \qquad \text{and} \qquad T_{2app} = \frac{T_0 - v_{02}X_0/C^2}{\sqrt{1 - v_{02}^2/C^2}} \quad (16)$$

Here $X_0$ and $T_0$ are the true co-ordinates of the vehicle relative to the system $S_1$ and these transformations can be regarded as Lorentz-Poincaré transformations. Only the interpretation of $X_{2app}$ and $T_{2app}$ differ from Poincare's approach.

### 3.3. Development of the comparison

We shall now further highlight the apparent similarities and the differences existing between conventional relativity and the aether theory presented here. Let us start from the expression of the extended space-time transformations relative to space (7).

*Space transformations:*
We have successively:

$$X_{1app} = X_{2app} \frac{\sqrt{1 - v_{02}^2/C^2}}{\sqrt{1 - v_{01}^2/C^2}} \frac{V - v_{01}}{V - v_{02}}$$

$$= X_{2app} \frac{(1-v_{02}^2/C^2)(V-v_{01})}{\sqrt{(1-v_{02}^2/C^2)(1-v_{01}^2/C^2)}(V-v_{02})}$$

$$= X_{2app} \frac{V-v_{01}-v_{02}^2 V/C^2 + v_{01}v_{02}^2/C^2}{(V-v_{02})\sqrt{1-v_{01}^2/C^2-v_{02}^2/C^2+v_{01}^2 v_{02}^2/C^4}}$$

$$= X_{2app} \frac{(V-v_{02})(1-v_{01}v_{02}/C^2)+(v_{02}-v_{01})(1-v_{02}V/C^2)}{(V-v_{02})\sqrt{(1-v_{01}v_{02}/C^2)^2 - \frac{(v_{02}-v_{01})^2}{C^2}}}$$

$$= X_{2app} \frac{\dfrac{(V-v_{02})(1-v_{01}v_{02}/C^2)+(v_{02}-v_{01})(1-v_{02}V/C^2)}{(V-v_{02})(1-v_{01}v_{02}/C^2)}}{\sqrt{\dfrac{C^2(1-v_{01}v_{02}/C^2)^2-(v_{02}-v_{01})^2}{C^2(1-v_{01}v_{02}/C^2)^2}}}$$

$$= \frac{X_{2app} + \dfrac{v_{02}-v_{01}}{1-v_{01}v_{02}/C^2} \dfrac{\dfrac{X_{2app}}{V-v_{02}}}{1-v_{02}V/C^2}}{\sqrt{1-\dfrac{(v_{02}-v_{01})^2}{C^2(1-v_{01}v_{02}/C^2)^2}}}$$

Finally :

$$X_{1app} = \frac{X_{2app}+\dfrac{v_{02}-v_{01}}{1-v_{01}v_{02}/C^2}T_{2app}}{\sqrt{1-\dfrac{(v_{02}-v_{01})^2}{C^2(1-v_{01}v_{02}/C^2)^2}}} \tag{17}$$

These transformations which relate the *apparent* distance and time measured in $S_1$ and $S_2$, differ from the Lorentz-Poincaré transformations. The conventional transformations, which do not recognize the existence of distorted co-ordinates, do not apply.

Notice that $v_{01}$ and $v_{02}$ are the true speeds that would be measured with non-contracted standards and with clocks not slowed down by motion and exactly synchronized.

When $S_1$ is at rest in the cosmic substratum, these transformations reduce to:

$$X_0 = \frac{X_{2app}+v_{02}T_{2app}}{\sqrt{1-v_{02}^2/C^2}}$$

And the reciprocal transformations take the form:

$$X_{2app} = \frac{X_0 - v_{02}T_0}{\sqrt{1 - v_{02}^2/C^2}}$$

a result which highlights the fact that the laws of nature are affected by the existence of the aether drift. (This result restricts the application of the relativity principle and the space-time transformations (17) do not constitute a group in all generality.) But when the speeds are measured with contracted standards and with clocks slowed down by motion and synchronized with light signals, their apparent value is equal to $v_{12app}$ such that :

$$v_{12app} = \frac{v_{02} - v_{01}}{1 - v_{01}v_{02}/C^2} \tag{18}$$

With these apparent speeds, the space transformations take *the same mathematical form* as the Lorentz-Poincaré transformations between any pair of 'inertial frames', their general form being:

$$X_{1app} = \frac{X_{2app} + v_{12app}T_{2app}}{\sqrt{1 - v_{12app}^2/C^2}} \quad \text{and} \quad X_{2app} = \frac{X_{1app} - v_{12app}T_{1app}}{\sqrt{1 - v_{12app}^2/C^2}}$$

and therefore the relativity principle seems to apply. Yet their meaning is quite different. Indeed obviously:

1/These transformations relate reference frames whose co-ordinates are altered by the measurement distortions, and therefore they give a distorted view of reality. Yet, these transformations are those which are obtained with the usual measurement procedures.

2/When $S_1$ is at rest in the Cosmic substratum, $v_{12app}$ reduces to $v_{02}$, $T_{1app}$ reduces to $T_0$ and $X_{1app}$ reduces to $X_0$. The physical reason of this fact is that in the aether frame, bodies are not submitted to the aether drift, while in the other frames they are affected by the drift, which entails length contraction, clock retardation and mass increase. Yet this result passes unnoticed. These special features differentiate aether theory from conventional relativity.

*Time transformations:*
We start from the expression of the extended space-time transformations relative to time (8). We have successively:

$$T_{1app} = T_{2app} \frac{\sqrt{1 - v_{02}^2/C^2}}{\sqrt{1 - v_{01}^2/C^2}} \frac{(1 - v_{01}V/C^2)}{(1 - v_{02}V/C^2)}$$

$$= T_{2app} \frac{(1-v_{02}^2/C^2)(1-v_{01}V/C^2)}{\sqrt{(1-v_{01}^2/C^2)(1-v_{02}^2/C^2)}(1-v_{02}V/C^2)}$$

$$= T_{2app} \frac{(1-v_{02}^2/C^2 - v_{01}V/C^2 + v_{01}v_{02}^2V/C^4)}{(1-v_{02}V/C^2)\sqrt{1-v_{01}^2/C^2 - v_{02}^2/C^2 + v_{01}^2v_{02}^2/C^4}}$$

$$= T_{2app} \frac{(1-v_{01}v_{02}/C^2)(1-v_{02}V/C^2) + \dfrac{(v_{02}-v_{01})(V-v_{02})}{C^2}}{(1-v_{02}V/C^2)\sqrt{1-v_{01}^2/C^2 - v_{02}^2/C^2 + v_{01}^2v_{02}^2/C^4}}$$

$$= T_{2app} \frac{(1-v_{01}v_{02}/C^2) + \dfrac{(v_{02}-v_{01})}{C^2}\dfrac{(V-v_{02})}{1-v_{02}V/C^2}}{\sqrt{1-v_{01}^2/C^2 - v_{02}^2/C^2 + 2v_{01}v_{02}/C^2 + v_{01}^2v_{02}^2/C^4 - 2v_{01}v_{02}/C^2}}$$

$$= \frac{\dfrac{T_{2app}(1-v_{01}v_{02}/C^2) + \dfrac{(v_{02}-v_{01})}{C^2}X_{2app}}{1-v_{01}v_{02}/C^2}}{\sqrt{\dfrac{(1-v_{01}v_{02}/C^2)^2 - \dfrac{(v_{01}-v_{02})^2}{C^2}}{(1-v_{01}v_{02}/C^2)^2}}}$$

Finally :

$$T_{1app} = \frac{T_{2app} + \dfrac{v_{02}-v_{01}}{1-v_{01}v_{02}/C^2}\dfrac{X_{2app}}{C^2}}{\sqrt{1-\dfrac{1}{C^2}\left(\dfrac{v_{02}-v_{01}}{1-v_{01}v_{02}/C^2}\right)^2}} \qquad (19)$$

The same reflections as those concerning the space transformations can be made when we replace the true speeds with the apparent speeds.

When $S_1$ is at rest in the Cosmic substratum, expression (19) reduces to:

$$T_0 = \frac{T_{2app} + v_{02}X_{2app}/C^2}{\sqrt{1-v_{02}^2/C^2}}$$

And the reciprocal transformation takes the form:

$$T_{2app} = \frac{T_0 - v_{02}X_0/C^2}{\sqrt{1-v_{02}^2/C^2}}$$

a result which highlights the fact that the laws of nature are affected by the existence of the aether drift. (This result restricts the application of the relativity principle, and the time transformations (19) do not constitute a group in all generality.)

But with the apparent speeds $v_{12app}$ measured with contracted standards and with clocks slowed down by motion and synchronized with light signals, the time transformations take *the same mathematical form* as the Lorentz-Poincaré transformations between any pair of 'inertial frames', their general form being:

$$T_{1app} = \frac{T_{2app} + v_{12app} X_{2app}/C^2}{\sqrt{1 - v_{12app}^2/C^2}} \quad \text{and} \quad T_{2app} = \frac{T_{1app} - v_{12app} X_{1app}/C^2}{\sqrt{1 - v_{12app}^2/C^2}}$$

and therefore the relativity principle seems to apply. Yet their meaning is quite different. Indeed in the same way as the space transformations:

1/These transformations relate reference frames whose co-ordinates are altered by the measurement distortions, and therefore they give a distorted view of reality. Yet these transformations are those which one obtains with the usual measurement procedures.

2/ When $S_1$ is at rest in the Cosmic substratum, $v_{12app}$ reduces to $v_{02}$, $T_{1app}$ reduces to $T_0$ and $X_{1app}$ reduces to $X_0$.

The same reasons for this fact as those invoked for the space transformations apply here. Yet this result passes unnoticed.

**Conclusion**

Although the mathematical form of the equations we have derived is identical to that of the Lorentz-Poincaré transformations, their meaning is completely different because they relate distorted co-ordinates and are dependent on an arbitrary synchronization procedure. Yet, these transformations are those which result from the experimental measurements.

As we have seen in formulas (15) and (16), they can be qualified as Lorentz-Poincaré transformations only when they connect the aether frame with any other 'inertial frame'.

The difference is all the more evident, as these transformations are derived from the extended space-time transformations, which assume the variability of the one-way speed of light when this speed is exactly measured, and show that the *apparent* invariance of the speed of light results from measurement distortions. If the synchronization were perfect, the speed of light would prove dependent on the relative speed between the fundamental frame and the frame where it is measured, a fact which would enable us to measure the absolute speed of this frame in contradiction with Poincaré's relativity principle [1]. And a near perfect clock synchronization is not a priori an objective impossible to reach.

It is clear that, if no preferred frame did exist, the celestial bodies, taken as reference systems would, in all probability, move in an almost absolute vacuum. In this case, the existence of near perfect inertial frames would be possible. Indeed, no physical effect could distinguish one frame from another. As a result, the laws of physics, relating exactly measured variables, would be identical in all these reference frames. But, as we demonstrated in ref [1], a number of experimental and theoretical arguments lend support to the existence of a preferred aether frame.

A preferred frame can only be conceived, if it is distinguished from the others in that a body at rest in it is submitted to distinct physical effects. This implies the existence of a medium, difficult to detect, but identifiable by its effects, that we refer to as the aether. The magnitude of the interaction of the medium with bodies at rest in a certain 'inertial frame' must therefore vary as a function of the relative speed between the frame considered and the aether frame. As the example of the two rockets demonstrates, *provided that the speeds are exactly measured*, the existence of the preferred aether frame proves incompatible with the exact application of the relativity principle. The space-time transformations we have derived, therefore, conceal hidden variables which deserved to be disclosed for a deeper understanding of physics.

**Post scriptum**
Although the present version provides additional explanations, the conclusions drawn do not differ from the previous version (physics/0607067, July 7[th] 2006).